\documentclass[a4paper,12pt]{article}

\usepackage[dvips]{graphicx}
\usepackage{geometry}
\usepackage[centertags]{amsmath}
\usepackage{amsfonts}
\usepackage{amssymb}
\usepackage{amsthm}
\usepackage{latexsym}
\usepackage{psfrag}

\numberwithin{equation}{section}
\renewcommand{\vec}[1]{\ensuremath{\boldsymbol{\mathrm{#1}}}}
\begin{document}

\sloppy

\title{\textbf{Speed dependent polarization correlations in QED and entanglement}%
\thanks{Work supported by the Royal Golden Jubilee Ph.D. Program.}~
\thanks{Published in \textit{European Physical Journal D}, Vol.~~\textbf{31}, No.~1, (2004)
pp~137--143.} }

\author{\textsc{E.~B.~Manoukian}\thanks{E-mail: \texttt{edouard@ccs.sut.ac.th}} \ and
\ \textsc{N.~Yongram} \\
{School of Physics, \ Suranaree University of Technology} \\
\ Nakhon Ratchasima, 30000, Thailand }
\date{} \maketitle

\begin{abstract}
Exact computations of polarizations correlations probabilities are
carried out in QED, to the leading order, for initially
\textit{polarized} as well as \textit{unpolarized} particles\@.
Quite generally they are found to be \textit{speed dependent} and
are in clear violation of Bell's inequality of Local Hidden
Variables (LHV) theories\@. This dynamical analysis shows how
speed dependent entangled states are generated\@. These
computations, based on QED are expected to lead to new experiments
on polarization correlations monitoring speed in the light of
Bell's theorem\@. The paper
provides a full QED treatment of the dynamics of entanglement\@.\\

quantum nonlocality (e.g. EPR paradox, Bells inequalities, GHZ
states, etc.) \\
\end{abstract}

\section{INTRODUCTION}\label{Sec1}
\noindent\indent We carry out exact computations of joint
probabilities of particle polarizations correlations in QED, to
the leading order, for initially \textit{polarized} and
\textit{unpolarized} particles\@. The interesting lesson we have
learnt from such studies is that the mere fact that particles
emerging from a process have non-zero speeds to reach detectors
implies, in general, that their polarizations correlations
probabilities \textit{depend} on speed \cite{B01}\@. The present
extended, and needless to say, dynamical analysis shows that this
is true, in general\@. This is unlike formal arguments based
simply on combining angular momenta\@. As a byproduct of this
work, we obtain clear violations with Bell's inequality (cf.
\cite{B02}--\cite{B04}) of LHV theories\@. We will also see how
QED generates speed
dependent entangled states\@.\\

Several experiments have been perfomed in recent years (cf.
\cite{B04}--\cite{B08}) on particles' polarizations
correlations\@. And, it is expected that the novel properties
recorded here by explicit calculations following directly from
field theory, which is based on the principle of relativity and
quantum theory, will lead to new experiments on polarization
correlations monitoring speed in the light of Bell's Theorem\@. We
hope that theses computations will be also useful in such areas of
physics
as quantum teleportation and quantum information in general\@.\\

The relevant quantity of interest here in testing Bell's
inequality of LHV \cite{B02} theories is, in a standard notation,

\begin{align}\label{Eqn1.1}
S&=\frac{p_{12}(a_1,a_2)}{p_{12}(\infty,\infty)}-\frac{p_{12}(a_1,a'_2)}{p_{12}(\infty,\infty)}
+\frac{p_{12}(a'_1,a_2)}{p_{12}(\infty,\infty)}+\frac{p_{12}(a'_1,a'_2)}{p_{12}(\infty,\infty)}
\nonumber\\[0.5\baselineskip]
&-\frac{p_{12}(a'_1,\infty)}{p_{12}(\infty,\infty)}-\frac{p_{12}(\infty,a_2)}{p_{12}(\infty,\infty)}
\end{align}
as is \textit{computed from} QED\@. Here $a_1$, $a_2\;(a'_1,a'_2)$
specify directions along which the polarizations of two particles
are measured, with $p_{12}(a_1,a_2)/p_{12}(\infty,\infty)$
denoting the joint probability, and
$p_{12}(a_1,\infty)/p_{12}(\infty,\infty)$,
$p_{12}(\infty,a_2)/p_{12}(\infty,\infty)$ denoting the
probabilities when the polarization of only one of the particles
is measured\@. [$p_{12}(\infty,\infty)$ is normalization
factor\@.] The corresponding probabilities as computed from QED
will be denoted by $P\left[\chi_1,\chi_2\right]$,
$P\left[\chi_1,-\right]$, $P\left[-,\chi_2\right]$ with $\chi_1$,
$\chi_2$ denoting angles the polarization vectors make with
certain axes spelled out in the bulk of the paper\@. To show that
QED is in violation with Bell's inequality of LHV, it is
sufficient to find one set of angles $\chi_1$, $\chi_2$,
$\chi'_1$, $\chi'_2$ and speed $\beta$, such that $S$, as computed
in QED, leads to a value of $S$ with $S>0$ or $S<-1$\@. In this
work, it is implicitly assumed that the polarization parameters in
the particle states are directly observable and may be used for
Bell-type measurements as discussed\@.\\

The need of a relativistic treatment based on explicit quantum
field \textit{dynamical} calculations in testing Bell-like
inequalities is critically important\@. An intriguing and very
recent reference \cite{B09}, which appeared after the submission
of our paper for publication, discusses the role of relativity in
quantum information, in general, and traces the historical
development of its role, and most importantly, in the light of our
present investigations, emphasizes the need of quantum field
theory as necessary for a consistent description of
interactions\@. Most earlier analyses dealing with relativistic
aspects, relevant to information theory and Bell-like tests are
kinematical of nature or deal with basic general properties of
local operators associated with bounded regions of spacetime
setting limits on measurements and localizability of quantum
systems\@. These probabilities are well documented in some of the
recent monographs \cite{B10}--\cite{B12} on the subject\@. Notable
important other recent references on such general aspects which
are, however, non-dynamical of nature are \cite{B13}--\cite{B17},
and a paper by Czachor \cite{B18} indicating how a possible
decrease in violation of Bell's inequalities may occur\@. In the
present work, we are interested in dynamical aspects and related
uniquely determined probabilities (intensities) of correlations
based on QED, as a fully relativistic quantum field theory (i.e.,
encompassing quantum theory and relativity) that meet the verdict
of experiments\@. QED is a non-speculative theory and as Feynman
\cite{B19} puts it, it is the most precise theory we have in
fundamental physics\@. The closest investigation to our own is
that of Ref.\cite{B20}, a reference we encountered after the
submission of our paper for publication, which considers spin-spin
interactions, in a QED setting, for non-relativistic electrons
and, unfortunately, does not compute their polarizations
correlations which are much relevant experimentally\@. In the
present paper, exact fully relativistic QED, computations, to the
leading order, of polarizations \textit{correlations} are
explicitly carried out for initially polarized and unpolarized
particles\@. The importance of also considering unpolarized spin
stems from the fact that we discover the existence of non-trivial
correlations, in the outcome of the processes, even for such mixed
states (since one averages over spin) and not only for pure states
arising from polarized spins, leading, in particular, in both
cases to speed dependent probabilities\@. The main results of our
paper are given in (\ref{Eqn2.20}), (\ref{Eqn2.22}),
(\ref{Eqn2.23}), (\ref{Eqn2.41})--(\ref{Eqn2.43}),
(\ref{Eqn3.10}), (\ref{Eqn3.12})--(\ref{Eqn3.19})\@. All of these
probabilities lead to a violation of Bell's inequality of LHV
theories\@. As the computations are based on the fully
relativistic QED, it is of some urgency that relevant experiments
are carried out by monitoring speed\@.

\section{POLARIZATIONS CORRELATIONS: INITIALLY POLARIZED PARTICLES}
\label{Sec2}

\noindent\indent We consider the process $e^-e^-\rightarrow
e^-e^-$, in the c.m., with initially polarized electrons with one
spin up, along the $z$-axis, and one spin down\@. With
$\vec{p}_1=\gamma m\beta(0,1,0)=-\vec{p}_2$ denoting the momenta
of the initial electrons, $\gamma=1/\sqrt{1-\beta^2}$, we consider
momenta of the emerging electrons with

\begin{equation}\label{Eqn2.1}
\vec{p}'_1=\gamma m\beta(\sin\theta,0,\cos\theta)=-\vec{p}'_2
\end{equation}
where $\theta$ is measured from the $z$-axis\@.\\

For the four-spinors of the initial electrons, we have
$(p^0=\gamma m)$

\begin{align}
  u(p_1)&=\left(\frac{p^0+m}{2m}\right)^{1/2}
  \begin{pmatrix}\begin{pmatrix}1 \\ 0\end{pmatrix}\\\\
  \mathrm{i}\rho\begin{pmatrix}0 \\ 1\end{pmatrix}
  \end{pmatrix}\label{Eqn2.2}\\[0.5\baselineskip]
  u(p_2)&=\left(\frac{p^0+m}{2m}\right)^{1/2}
  \begin{pmatrix}\begin{pmatrix}0 \\ 1\end{pmatrix}\\\\
  \mathrm{i}\rho\begin{pmatrix}1 \\ 0\end{pmatrix}
  \end{pmatrix}\label{Eqn2.3}\\[0.5\baselineskip]
  \rho&=\frac{\gamma\beta}{\gamma+1}=\frac{\beta}{1+\sqrt{1-\beta^2}}
  \label{Eqn2.4}
\end{align}
and for the final ones

\begin{align}
  u(p'_1)&=\left(\frac{p^0+m}{2m}\right)^{1/2}
  \begin{pmatrix}\xi_1\\
  \frac{\vec{\sigma}\cdot\vec{p}'_1}{p^0+m}\xi_1
  \end{pmatrix}\label{Eqn2.5}\\[0.5\baselineskip]
  u(p'_2)&=\left(\frac{p^0+m}{2m}\right)^{1/2}
  \begin{pmatrix}\xi_2 \\
  -\frac{\vec{\sigma}\cdot\vec{p}'_1}{p^0+m}\xi_2
  \end{pmatrix}\label{Eqn2.6}
\end{align}
where the two-spinors $\xi_1$, $\xi_2$ will be specified
later\@.\\

The expression for the amplitude of the process is well known (cf.
\cite{B21})

\begin{eqnarray}\label{Eqn2.7}
  A\propto\left[\frac{\overline{u}(p'_1)\gamma^\mu u(p_1)
  \overline{u}(p'_2)\gamma_\mu u(p_2)}{(p'_1-p_1)^2}-\frac{\overline{u}(p'_2)\gamma^\mu u(p'_1)
  \overline{u}(p'_1)\gamma_\mu u(p_2)}{(p'_2-p_1)^2}\right]
\end{eqnarray}

The following matrix elements are needed and are readily
calculated

\begin{align}
  \overline{u}(p'_1)\gamma^0u(p_1)&=\frac{p^0+m}{2m}\xi^\dag_1
  \begin{pmatrix}1+\mathrm{i}\rho^2\sin\theta\\
  -\mathrm{i}\rho^2\cos\theta
  \end{pmatrix}\label{Eqn2.8}\\[0.5\baselineskip]
  \overline{u}(p'_2)\gamma^0u(p_2)&=\frac{p^0+m}{2m}\xi^\dag_2
  \begin{pmatrix}-\mathrm{i}\rho^2\cos\theta\\
  1-\mathrm{i}\rho^2\sin\theta
  \end{pmatrix}\label{Eqn2.9}\\[0.5\baselineskip]
  \overline{u}(p'_1)\gamma^0u(p_2)&=\frac{p^0+m}{2m}\xi^\dag_1
  \begin{pmatrix}\mathrm{i}\rho^2\cos\theta\\
  1+\mathrm{i}\rho^2\sin\theta
  \end{pmatrix}\label{Eqn2.10}\\[0.5\baselineskip]
  \overline{u}(p'_2)\gamma^0u(p_1)&=\frac{p^0+m}{2m}\xi^\dag_2
  \begin{pmatrix}1-\mathrm{i}\rho^2\sin\theta\\
  \mathrm{i}\rho^2\cos\theta
  \end{pmatrix}\label{Eqn2.11}\\[0.5\baselineskip]
  \overline{u}(p'_1)\gamma^ju(p_1)&=\frac{p^0+m}{2m}\rho\xi^\dag_1
  \bigg[\begin{pmatrix}\mathrm{i}+\sin\theta \\ -\cos\theta\end{pmatrix}\delta^{j1}
  +\mathrm{i}\begin{pmatrix}-\mathrm{i}+\sin\theta\\-\cos\theta
  \end{pmatrix}\delta^{j2}+\begin{pmatrix}-\cos\theta\\
  -\mathrm{i}+\sin\theta
  \end{pmatrix}\delta^{j3}\bigg]\label{Eqn2.12}\\[0.5\baselineskip]
  \overline{u}(p'_2)\gamma^ju(p_2)&=\frac{p^0+m}{2m}\rho\xi^\dag_2
  \bigg[\begin{pmatrix}-\cos\theta \\ \mathrm{i}-\sin\theta\end{pmatrix}\delta^{j1}
  +\mathrm{i}\begin{pmatrix}\cos\theta\\ \mathrm{i}+\sin\theta
  \end{pmatrix}\delta^{j2}+\begin{pmatrix}\mathrm{i}+\sin\theta\\
  -\cos\theta
  \end{pmatrix}\delta^{j3}\bigg]\label{Eqn2.13}\\[0.5\baselineskip]
  \overline{u}(p'_1)\gamma^ju(p_2)&=\frac{p^0+m}{2m}\rho\xi^\dag_1
  \bigg[\begin{pmatrix}\cos\theta \\ \mathrm{i}+\sin\theta\end{pmatrix}\delta^{j1}
  +\mathrm{i}\begin{pmatrix}-\cos\theta\\ \mathrm{i}-\sin\theta
  \end{pmatrix}\delta^{j2}+\begin{pmatrix}\mathrm{i}-\sin\theta\\
  \cos\theta
  \end{pmatrix}\delta^{j3}\bigg]\label{Eqn2.14}\\[0.5\baselineskip]
  \overline{u}(p'_2)\gamma^ju(p_1)&=\frac{p^0+m}{2m}\rho\xi^\dag_2
  \bigg[\begin{pmatrix}\mathrm{i}-\sin\theta \\ \cos\theta\end{pmatrix}\delta^{j1}
  -\mathrm{i}\begin{pmatrix}\mathrm{i}+\sin\theta\\ -\cos\theta
  \end{pmatrix}\delta^{j2}-\begin{pmatrix}\cos\theta\\
  \mathrm{i}+\sin\theta
  \end{pmatrix}\delta^{j3}\bigg]\label{Eqn2.15}
\end{align}

\begin{figure}[h!]
  \centering
  \fbox{ \begin{minipage}{0.9\textwidth}
  \centering
  \psfrag{x}{$x$}
  \psfrag{z}{$z$}
  \psfrag{y}{$y$}
  \psfrag{c}{$\chi_1$}
  \psfrag{e}{$e^-$}
  \includegraphics[width=0.8\textwidth]{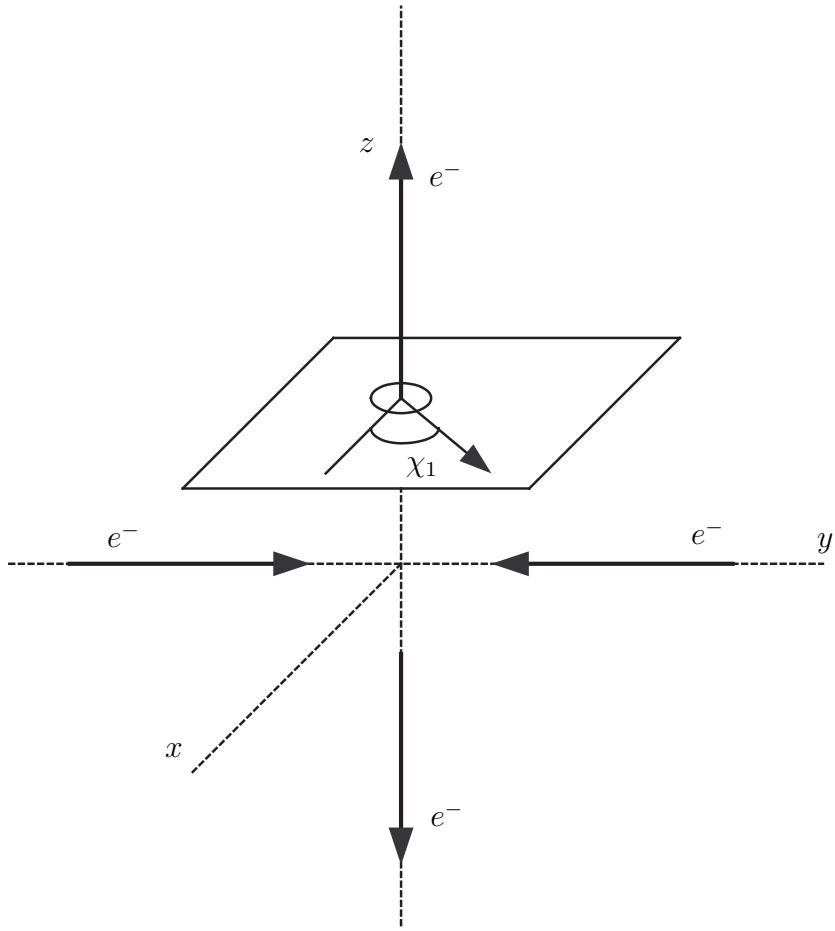}
  \caption{The figure depicts $e^-e^-$ scattering, with the electrons
   initially moving along the $y$-axis, while the emerging electrons
   moving along the $z$-axis\@. The angle $\chi_1$, measured relative to
   the $x$-axis, denotes the orientation of spin of one of the emerging
   electrons may make\@.}
           \label{Fig1}
   \end{minipage} }
\end{figure}

For $\theta=0$, (see FIG.~\ref{Fig1}), we obtain from
(\ref{Eqn2.7})--(\ref{Eqn2.15})

\begin{align}\label{Eqn2.16}
  A\propto
  \xi^\dag_1\xi^\dag_2\Bigg\{(1+6\rho^2+\rho^4)&\left[
  \begin{pmatrix}0 \\1\end{pmatrix}_1\begin{pmatrix}1 \\0\end{pmatrix}_2
  -\begin{pmatrix}1 \\0\end{pmatrix}_1\begin{pmatrix}0 \\1\end{pmatrix}_2\right]\nonumber
  \\[0.5\baselineskip]
  &+4\mathrm{i}\rho^2\left[
  \begin{pmatrix}0 \\1\end{pmatrix}_1\begin{pmatrix}0 \\1\end{pmatrix}_2
  +\begin{pmatrix}1 \\0\end{pmatrix}_1\begin{pmatrix}1
  \\0\end{pmatrix}_2\right]\Bigg\}
\end{align}
generating the speed dependent (normalized) entangled state of the
emerging electrons

\begin{align}\label{Eqn2.17}
  \left|\psi\right>&=\frac{1}{N}\Bigg\{\frac{(1+6\rho^2+\rho^4)}{\sqrt{2}}\left[
  \begin{pmatrix}0 \\1\end{pmatrix}_1\begin{pmatrix}1 \\0\end{pmatrix}_2
  -\begin{pmatrix}1 \\0\end{pmatrix}_1\begin{pmatrix}0 \\1\end{pmatrix}_2\right]\nonumber
  \\[0.5\baselineskip]
  &+\frac{4\mathrm{i}\rho^2}{\sqrt{2}}\left[
  \begin{pmatrix}0 \\1\end{pmatrix}_1\begin{pmatrix}0 \\1\end{pmatrix}_2
  +\begin{pmatrix}1 \\0\end{pmatrix}_1\begin{pmatrix}1
  \\0\end{pmatrix}_2\right]\Bigg\}
\end{align}
where

\begin{align}
  N&=\left[(1+6\rho^2+\rho^4)^2+16\rho^4\right]^{1/2}\label{Eqn2.18}\\[0.5\baselineskip]
  \xi_j&=\frac{1}{\sqrt{2}}\begin{pmatrix}\mathrm{e}^{-\mathrm{i}\chi_j/2}
  \\[0.5\baselineskip]
  \mathrm{e}^{\mathrm{i}\chi_j/2}\end{pmatrix},\quad j=1,2\label{Eqn2.19}
\end{align}
$\rho$ is defined in ($\ref{Eqn2.4}$), and the angles are measured
relative to the $x$-axis (see FIG.~\ref{Fig1})\@.\\

The joint probability of the electrons polarizations correlations
is then given by

\begin{align}\label{Eqn2.20}
  P[\chi_1,\chi_2]&=\left\|\xi^\dag_1\xi^\dag_2\left|\psi\right>\right\|^2
  \nonumber\\[0.5\baselineskip]
  &=\frac{\left[(1+6\rho^2+\rho^4)\sin\left(\frac{\chi_1-\chi_2}{2}\right)
  -4\rho^2\cos\left(\frac{\chi_1+\chi_2}{2}\right)\right]^2}
  {2\left[(1+6\rho^2+\rho^4)^2+16\rho^4\right]}
\end{align}
[For $\beta\rightarrow 0$, one obtains a rather familiar
expression
$P[\chi_1,\chi_2]=\sin^2\left[(\chi_1-\chi_2)/2\right]/2$\@.]\\

If only one of the spins is measured, say, corresponding to
$\chi_1$, we then have to form the state

\begin{align}\label{Eqn2.21}
  \xi^\dag_1\left|\psi\right>=\frac{[1+6\rho^2+\rho^4]}{2N}
  \begin{pmatrix}\mathrm{e}^{-\mathrm{i}\chi_1/2}\\
  -\mathrm{e}^{\mathrm{i}\chi_1/2}\end{pmatrix}_2+\frac{4\mathrm{i}\rho^2}{2N}
  \begin{pmatrix}\mathrm{e}^{\mathrm{i}\chi_1/2}\\
  \mathrm{e}^{-\mathrm{i}\chi_1/2}\end{pmatrix}_2
\end{align}
from which we obtain the corresponding probability

\begin{align}\label{Eqn2.22}
  P[\chi_1,-]&=\left\|\xi^\dag_1\left|\psi\right>\right\|^2
  \nonumber\\[0.5\baselineskip]
  &=\frac{1}{2}-\frac{4\rho^2(1+6\rho^2+\rho^4)}
  {(1+6\rho^2+\rho^4)^2+16\rho^4}\sin\chi_1
\end{align}
and similarly

\begin{align}\label{Eqn2.23}
  P[-,\chi_2]&=\left\|\xi^\dag_2\left|\psi\right>\right\|^2
  \nonumber\\[0.5\baselineskip]
  &=\frac{1}{2}+\frac{4\rho^2(1+6\rho^2+\rho^4)}
  {(1+6\rho^2+\rho^4)^2+16\rho^4}\sin\chi_2
\end{align}

The probability $P[\chi_1,-]$ may be \textit{equivalently}
obtained by summing $P[\chi_1,\chi_2]$ over the two angles

\begin{align}\label{Eqn2.24}
\chi_2,\quad \chi_2+\pi
\end{align}
for any arbitrarily chosen fixed $\chi_2$, i.e.,

\begin{align}\label{Eqn2.25}
P[\chi_1,\chi_2]+P[\chi_1,\chi_2+\pi]=P[\chi_1,-]
\end{align}
as is easily checked, and similarly for $P[-,\chi_2]$\@.\\

For all $0\leqslant\beta\leqslant1$, angles $\chi_1$, $\chi_2$,
$\chi'_1$, $\chi'_2$ are readily found leading to a violation of
Bell's inequality of LHV theories\@. For example, for $\beta=0.3$,
$\chi_1=0^{\circ}$, $\chi_2=137^{\circ}$, $\chi'_1=12^{\circ}$,
$\chi'_2=45^{\circ}$, $S=-1.79$ violating the inequality from
below\@.\\

The speed dependence of $P[\chi_1,\chi_2]$ generally holds true
for other angles as well\@. For $\theta=\pi/2$, however, it is
readily verified that (\ref{Eqn2.7}) leads to the entangled state

\begin{align}\label{Eqn2.26}
  \left|\psi\right>_0=\frac{1}{\sqrt{2}}\left[
  \begin{pmatrix}0 \\1\end{pmatrix}_1\begin{pmatrix}1 \\0\end{pmatrix}_2
  -\begin{pmatrix}1 \\0\end{pmatrix}_1\begin{pmatrix}0 \\1\end{pmatrix}_2\right]
\end{align}
for all $0\leqslant\beta\leqslant1$, leading to a rather familiar
expression
$P[\chi_1,\chi_2]=\sin^2\left[(\chi_1-\chi_2)/2\right]/2$\@.\\

Now we consider the process $e^+e^-\rightarrow2\gamma$, in the
c.m. of $e^-$, $e^+$ with spins up, along the $z$-axis, and down,
respectively\@. With $\vec{p}_1=\vec{p}(e^-)=\gamma
m\beta(0,1,0)=-\vec{p}(e^+)=-\vec{p}_2$, we have for $e^-$, $e^+$
the spinors given by

\begin{align}
  u&=\left(\frac{p^0+m}{2m}\right)^{1/2}
  \begin{pmatrix}\begin{pmatrix}1 \\ 0\end{pmatrix}\\\\
  \mathrm{i}\rho\begin{pmatrix}0 \\ 1\end{pmatrix}
  \end{pmatrix}\label{Eqn2.27}\\[0.5\baselineskip]
  v&=\left(\frac{p^0+m}{2m}\right)^{1/2}
  \begin{pmatrix}\mathrm{i}\rho\begin{pmatrix}0 \\ 1\end{pmatrix}\\\\
  \begin{pmatrix}1 \\ 0\end{pmatrix}
  \end{pmatrix}\label{Eqn2.28}
\end{align}
with $\rho$ defined in (\ref{Eqn2.4}), and we consider momenta of
the photons

\begin{align}\label{Eqn2.29}
\vec{k}_1=\gamma m(\sin\theta,0,\cos\theta)=-\vec{k}_2
\end{align}
where we have used the facts that

\begin{align}\label{Eqn2.30}
|\vec{k}_1|=|\vec{k}_2|=k^0_1=k^0_2=p^0(e^{\pm})\equiv p^0=\gamma
m
\end{align}

The amplitude for the process is given by (cf.\cite{B21})

\begin{align}\label{Eqn2.31}
A\propto\overline{v}\left[\frac{\gamma^\mu\gamma
k_1\gamma^\nu}{2p_1k_1}+\frac{\gamma^\nu\gamma
k_2\gamma^\mu}{2p_1k_2}+\frac{\gamma^\mu
p^\nu_1}{p_1k_1}+\frac{\gamma^\nu p^\mu_1}{p_1k_2}\right]u\;
e^\nu_1e^\mu_2
\end{align}
where $e^\mu_1=(0,\vec{e}_1)$, $e^\mu_2=(0,\vec{e}_2)$ are the
polarizations of the photons with $(j=1,2)$

\begin{align}\label{Eqn2.32}
\vec{e}_j=(-\cos\theta\cos\chi_j,\sin\chi_j,\sin\theta\cos\chi_j)\equiv
(e^{(1)}_j,e^{(2)}_j,e^{(3)}_j)
\end{align}

The following matrix elements are readily derived

\begin{align}
  \overline{v}\left(\gamma^i\gamma^0\gamma^j\right)u&=\frac{p^0+m}{2m}
  2\mathrm{i}\varepsilon_{ij2}\rho\label{Eqn2.33}\\[0.5\baselineskip]
  \overline{v}\gamma^iu&=\frac{p^0+m}{2m}
  (1-\rho^2)\delta^{i3}\label{Eqn2.34}\\[0.5\baselineskip]
  \overline{v}\left(\gamma^i\gamma^m\gamma^j\right)u&=\frac{p^0+m}{2m}
  \left(-\delta^{mj}\delta^{i3}-\delta^{mi}\delta^{j3}+
  \delta^{ij}\delta^{m3}\right)(1-\rho^2)\nonumber\\[0.5\baselineskip]
  &-\mathrm{i}\frac{p^0+m}{2m}
  (1+\rho^2)\varepsilon_{imj}\label{Eqn2.35}
\end{align}

Upon setting,

\begin{align}\label{Eqn2.36}
\frac{\vec{k}_1}{|\vec{k}_1|}=\vec{n}
\end{align}
the amplitude $A$ is then given by

\begin{align}\label{Eqn2.37}
A\propto-\mathrm{i}(1+\rho^2)\vec{n}\cdot(\vec{e}_1\times\vec{e}_2)
+\beta(1-\rho^2)\left(e^{(2)}_1e^{(3)}_2+e^{(3)}_1e^{(2)}_2\right)
\end{align}

\begin{figure}[h!]
  \centering
  \fbox{ \begin{minipage}{0.9\textwidth}
  \centering
  \psfrag{x}{$x$}
  \psfrag{z}{$z$}
  \psfrag{y}{$y$}
  \psfrag{c}{$\chi_1$}
  \psfrag{e-}{$e^-$}
  \psfrag{e+}{$e^+$}
  \includegraphics[width=0.8\textwidth]{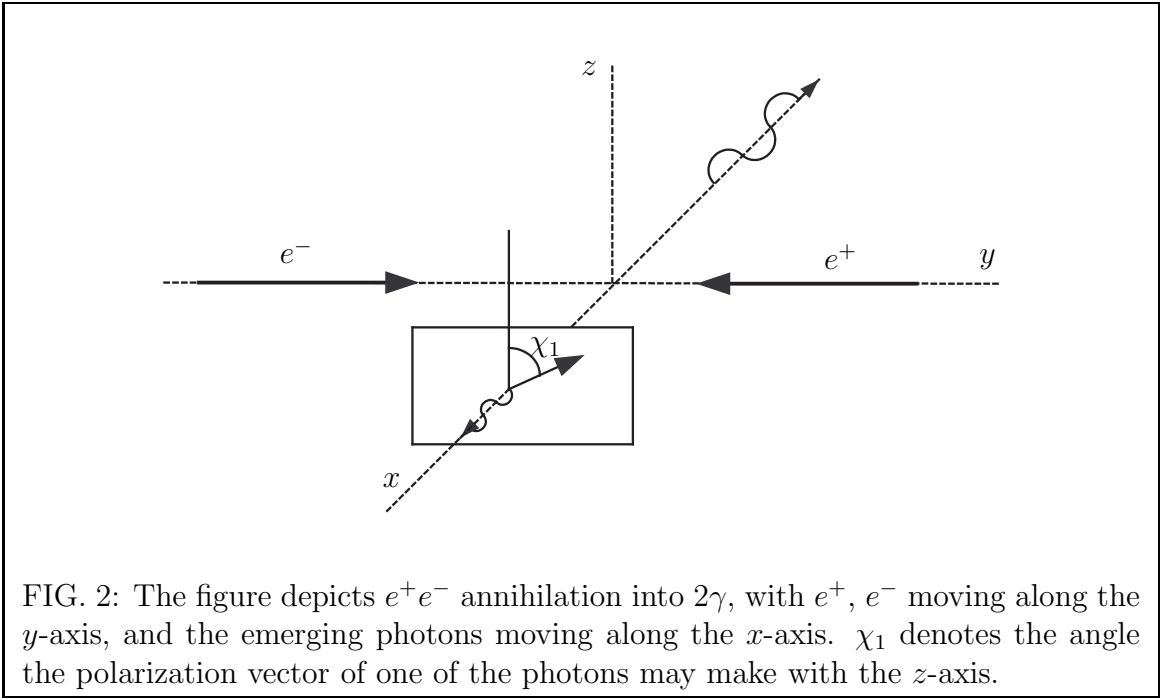}
  \caption{The figure depicts $e^+e^-$ annihilation into $2\gamma$, with $e^+$, $e^-$
   moving along the $y$-axis, and the emerging photons
   moving along the $x$-axis\@. $\chi_1$ denotes the angle the polarization vector of one
   of the photons may make with the $z$-axis\@.}
           \label{Fig2}
   \end{minipage} }
\end{figure}

For $\theta=\pi/2$, this gives

\begin{align}\label{Eqn2.38}
  A&\propto-\left(0,\sin\chi_1,\sin\chi_1\right)_1\left(0,\sin\chi_2,\sin\chi_2\right)_2
  \nonumber\\[0.5\baselineskip]
  &\times\Bigg\{\mathrm{i}(1+\rho^2)\left[\begin{pmatrix}0 \\1 \\
  0\end{pmatrix}_1\begin{pmatrix}0
  \\0\\1\end{pmatrix}_2
  -\begin{pmatrix}0 \\0 \\ 1\end{pmatrix}_1\begin{pmatrix}0 \\1 \\ 0\end{pmatrix}_2\right]\nonumber
  \\[0.5\baselineskip]
  &-\beta(1-\rho^2)\left[
  \begin{pmatrix}0 \\1 \\ 0\end{pmatrix}_1\begin{pmatrix}0\\ 0 \\1\end{pmatrix}_2
  +\begin{pmatrix}0 \\0 \\ 1\end{pmatrix}_1\begin{pmatrix}0\\ 1
  \\0\end{pmatrix}_2\right]\Bigg\}
\end{align}
(see FIG.~\ref{Fig2}), generating a speed dependent (normalized)
entangled state for the photons given by

\begin{align}\label{Eqn2.39}
 \left|\phi\right>&=\frac{1}{N}\Bigg\{\frac{\mathrm{i}(1+\rho^2)}{\sqrt{2}}\left[\begin{pmatrix}0 \\1 \\
  0\end{pmatrix}_1\begin{pmatrix}0
  \\0\\1\end{pmatrix}_2
  -\begin{pmatrix}0 \\0 \\ 1\end{pmatrix}_1\begin{pmatrix}0 \\1 \\ 0\end{pmatrix}_2\right]\nonumber
  \\[0.5\baselineskip]
  &-\beta\frac{(1-\rho^2)}{\sqrt{2}}\left[
  \begin{pmatrix}0 \\1 \\ 0\end{pmatrix}_1\begin{pmatrix}0\\ 0 \\1\end{pmatrix}_2
  +\begin{pmatrix}0 \\0 \\ 1\end{pmatrix}_1\begin{pmatrix}0\\ 1
  \\0\end{pmatrix}_2\right]\Bigg\}
\end{align}
with

\begin{align}\label{Eqn2.40}
N=\left[(1+\rho^2)^2+\beta^2(1-\rho^2)^2\right]^{1/2}
\end{align}

Therefore the joint probability of photons polarizations
correlations is given by

\begin{align}\label{Eqn2.41}
  P[\chi_1,\chi_2]&=\left\|(0,\sin\chi_1,\cos\chi_1)_1(0,\sin\chi_2,\cos\chi_2)_2\left|\phi\right>\right\|^2
  \nonumber\\[0.5\baselineskip]
  &=\frac{(1+\rho^2)^2\sin^2(\chi_1-\chi_2)+\beta^2(1-\rho^2)^2\cos^2(\chi_1+\chi_2)}
  {2[(1+\rho^2)^2+\beta^2(1-\rho^2)^2]}
\end{align}
and

\begin{align}
  P[\chi_1,-]&=\left\|(0,\sin\chi_1,\cos\chi_1)_1\left|\phi\right>\right\|^2
  =\frac{1}{2}\label{Eqn2.42}\\[0.5\baselineskip]
  P[-,\chi_2]&=\left\|(0,\sin\chi_2,\cos\chi_2)_2\left|\phi\right>\right\|^2
  =\frac{1}{2}\label{Eqn2.43}
\end{align}
$P[\chi_1,-]$ is also \textit{equivalently} obtained by summing
$P[\chi_1,\chi_2]$ over

\begin{align}\label{Eqn2.44}
\chi_2,\quad \chi_2+\frac{\pi}{2}
\end{align}
for any arbitrarily chosen $\chi_2$, i.e.,

\begin{align}\label{Eqn2.45}
P[\chi_1,\chi_2]+P[\chi_1,\chi_2+\frac{\pi}{2}]=P[\chi_1,-]
\end{align}
and similarly for $P[-,\chi_2]$\@.\\

For all $0\leqslant\beta\leqslant1$, angles $\chi_1$, $\chi_2$,
$\chi'_1$, $\chi'_2$ are readily found leading to a violation of
Bell's inequality of LHV theories\@. For example, for $\beta=0.2$,
$\chi_1=0^{\circ}$, $\chi_2=23^{\circ}$, $\chi'_1=45^{\circ}$,
$\chi'_2=67^{\circ}$, $S=-1.187$ violating the inequality from
below\@.\\

Again the speed dependence of $P[\chi_1,\chi_2]$ generally holds
true for other angles as well\@. For $\theta=0$, however, it is
readily checked that (\ref{Eqn2.37}) leads to the entangled state

\begin{align}\label{Eqn2.46}
  \left|\phi_0\right>=\frac{1}{\sqrt{2}}\left[
  \begin{pmatrix}0 \\1 \\0\end{pmatrix}_1\begin{pmatrix}0 \\0 \\1\end{pmatrix}_2
  -\begin{pmatrix}0\\0\\1\end{pmatrix}_1\begin{pmatrix}0 \\1\\0\end{pmatrix}_2\right]
\end{align}
for all $0\leqslant\beta\leqslant1$ giving a rather familiar
expression
$P[\chi_1,\chi_2]=\left[\sin^2(\chi_1-\chi_2)\right]/2$\@.

\section{POLARIZATIONS CORRELATIONS: INITIALLY UNPOLARIZED
PARTICLES}\label{Sec3}

\noindent\indent For the process $e^-e^-\rightarrow e^-e^-$, in
the c.m., with initially unpolarized spins, with momenta
$\vec{p}_1=\gamma m\beta(0,1,0)=-\vec{p}_2$, we take for the final
electrons

\begin{align}\label{Eqn3.1}
 \vec{p'}_1=\gamma m\beta(1,0,0)=-\vec{p}'_2
\end{align}
and for the four-spinors

\begin{align}
  u(p'_1)&=\left(\frac{p^0+m}{2m}\right)^{1/2}
  \begin{pmatrix}\xi_1\\
  \frac{\vec{\sigma}\cdot\vec{p}'_1}{p^0+m}\xi_1
  \end{pmatrix},\quad \xi_1=\begin{pmatrix}-\mathrm{i}\cos\chi_1/2\\
  \sin\chi_1/2
  \end{pmatrix}\label{Eqn3.2}\\[0.5\baselineskip]
  u(p'_2)&=\left(\frac{p^0+m}{2m}\right)^{1/2}
  \begin{pmatrix}\xi_2 \\
  -\frac{\vec{\sigma}\cdot\vec{p}'_1}{p^0+m}\xi_2
  \end{pmatrix},\quad \xi_2=\begin{pmatrix}-\mathrm{i}\cos\chi_2/2\\
  \sin\chi_2/2\end{pmatrix}\label{Eqn3.3}
\end{align}
A straightforward but tedious computation of the corresponding
probability of occurrence with initially unpolarized electrons,
(\ref{Eqn2.7}) leads to

\begin{align}\label{Eqn3.4}
 \text{Prob}&\propto\left[\overline{u}(p'_1)\gamma^\mu(-\gamma p_1+m)\gamma^\sigma u(p'_1)\right]
 \left[\overline{u}(p'_2)\gamma_\mu(-\gamma p_2+m)\gamma_\sigma u(p'_2)\right]
 \nonumber\\[0.5\baselineskip]
 &-\left[\overline{u}(p'_1)\gamma^\mu(-\gamma p_1+m)\gamma^\sigma u(p'_2)\right]
 \left[\overline{u}(p'_2)\gamma_\mu(-\gamma p_2+m)\gamma_\sigma u(p'_1)\right]
 \nonumber\\[0.5\baselineskip]
 &-\left[\overline{u}(p'_2)\gamma^\mu(-\gamma p_1+m)\gamma^\sigma u(p'_1)\right]
 \left[\overline{u}(p'_1)\gamma_\mu(-\gamma p_2+m)\gamma_\sigma u(p'_2)\right]
 \nonumber\\[0.5\baselineskip]
 &+\left[\overline{u}(p'_2)\gamma^\mu(-\gamma p_1+m)\gamma^\sigma u(p'_2)\right]
 \left[\overline{u}(p'_1)\gamma_\mu(-\gamma p_2+m)\gamma_\sigma u(p'_1)\right]
\end{align}
which after simplification and of collecting terms reduces to

\begin{align}\label{Eqn3.5}
 \text{Prob}&\propto(1-\beta^2)(1+3\beta^2)\sin^2\left(\frac{\chi_1-\chi_2}{2}\right)
 +\beta^4\cos^2\left(\frac{\chi_1+\chi_2}{2}\right)+4\beta^4\nonumber\\[0.5\baselineskip]
 &\equiv F[\chi_1,\chi_2]
\end{align}
where we have used the expressions for the spinors in
(\ref{Eqn3.2}), (\ref{Eqn3.3})\@.\\

Given that the process has occurred, the conditional probability
that the spins of the emerging electrons make angles $\chi_1$,
$\chi_2$ with the $z$-axis, is directly obtained from
(\ref{Eqn3.5}) to be

\begin{align}\label{Eqn3.6}
 P[\chi_1,\chi_2]=\frac{F[\chi_1,\chi_2]}{C}
\end{align}
The normalization constant $C$ is obtained by summing over the
polarizations of the emerging electrons\@. This is equivalent to
summing of $F[\chi_1,\chi_2]$ over the pairs of angles

\begin{align}\label{Eqn3.7}
 (\chi_1,\chi_2),\;(\chi_1+\pi,\chi_2),\;(\chi_1,\chi_2+\pi)\;(\chi_1+\pi,\chi_2+\pi)
\end{align}
for any arbitrarily chosen fixed $\chi_1$, $\chi_2$, corresponding
to the orthonormal spinors

\begin{align}\label{Eqn3.8}
 \begin{pmatrix}-\mathrm{i}\cos\chi_j/2\\
  \sin\chi_j/2\end{pmatrix},\quad\begin{pmatrix}-\mathrm{i}\cos(\chi_j+\pi)/2\\
  \sin(\chi_j+\pi)/2\end{pmatrix}=\begin{pmatrix}\mathrm{i}\sin\chi_j/2\\
  \cos\chi_j/2\end{pmatrix}
\end{align}
providing a complete set, for each $j=1,2$, in reference to
(\ref{Eqn3.2}), (\ref{Eqn3.3})\@.This is,

\begin{align}\label{Eqn3.9}
 C&=F[\chi_1,\chi_2]+F[\chi_1+\pi,\chi_2]+F[\chi_1,\chi_2+\pi]+F[\chi_1+\pi,\chi_2+\pi]
 \nonumber\\[0.5\baselineskip]
 &=2(1+2\beta^2+6\beta^4)
\end{align}
which as expected is independent of $\chi_1$, $\chi_2$, giving

\begin{align}\label{Eqn3.10}
 P[\chi_1,\chi_2]=\frac{(1-\beta^2)(1+3\beta^2)\sin^2\left(\frac{\chi_1-\chi_2}{2}\right)
 +\beta^4\cos^2\left(\frac{\chi_1+\chi_2}{2}\right)+4\beta^4}{2(1+2\beta^2+6\beta^4)}
\end{align}

By summing over

\begin{align}\label{Eqn3.11}
\chi_2,\quad \chi_2+\pi
\end{align}
for any arbitrarily fixed $\chi_2$, we obtain

\begin{align}\label{Eqn3.12}
P[\chi_1,-]=\frac{1}{2}
\end{align}
and similarly,

\begin{align}\label{Eqn3.13}
P[-,\chi_2]=\frac{1}{2}
\end{align}
for the probabilities when only one of the photons polarizations
is measured\@.\\

A clear violation of Bell's inequality of LHV theories was
obtained for all $0\leqslant\beta\leqslant0.45$\@. For example,
for $\beta=0.3$, with $\chi_1=0^{\circ}$, $\chi_2=45^{\circ}$,
$\chi'_1=90^{\circ}$, $\chi'_2=135^{\circ}$ give $S=-1.165$
violating the inequality from below\@. For larger $\beta$ values,
alone, one cannot discriminate between LHV theories and quantum
theory for this process\@. A violation of Bell's inequality for at
least some $\beta$ values, as seen, however, automatically
violates LHV theories\@.\\

The probability of photon polarizations correlations in
$e^+e^-\rightarrow2\gamma$ with initially unpolarized $e^+$,
$e^-$, has been given in \cite{B01} to be

\begin{align}
P[\chi_1,\chi_2]&=
\frac{1-\left[\cos(\chi_1-\chi_2)-2\beta^2\cos\chi_1\cos\chi_2\right]^2}
{2[1+2\beta^2(1-\beta^2)]}\label{Eqn3.14}\\[0.5\baselineskip]
P[\chi_1,-]&= \frac{1+4\beta^2(1-\beta^2)\cos^2\chi_1}
{2[1+2\beta^2(1-\beta^2)]}\label{Eqn3.15}\\[0.5\baselineskip]
P[-,\chi_2]&= \frac{1+4\beta^2(1-\beta^2)\cos^2\chi_2}
{2[1+2\beta^2(1-\beta^2)]}\label{Eqn3.16}
\end{align}
and a clear violation of Bell's inequality of LHV theories was
obtained for all $0\leqslant\beta\leqslant0.2$\@. Again, for
larger values of $\beta$, alone, one cannot discriminate between
LHV theories and quantum theory for this process\@. A violation of
Bell's inequality for at least some $\beta$ values, as seen,
however, automatically occurs violating LHV theories\@.\\

For completeness, we mention that for the annihilation of the
spin~$0$ pair into $2\gamma$ the following probabilities are
similarly worked out:

\begin{align}
P[\chi_1,\chi_2]&=
\frac{\left(\cos(\chi_1-\chi_2)-2\beta^2\cos\chi_1\cos\chi_2\right)^2}
{2[1-2\beta^2(1-\beta^2)]}\label{Eqn3.17}\\[0.5\baselineskip]
P[\chi_1,-]&= \frac{1-4\beta^2(1-\beta^2)\cos^2\chi_1}
{2[1-2\beta^2(1-\beta^2)]}\label{Eqn3.18}\\[0.5\baselineskip]
P[-,\chi_2]&= \frac{1-4\beta^2(1-\beta^2)\cos^2\chi_2}
{2[1-2\beta^2(1-\beta^2)]}\label{Eqn3.19}
\end{align}
and violates Bell's inequality of LHV theories for all
$0\leqslant\beta\leqslant1$\@.

\section{CONCLUSION}\label{Sec4}

\noindent\indent We have seen by explicit dynamical computations
based on QED, that the polarizations correlations probabilities of
particles emerging in processes \textit{depend} on speed, for
initially \textit{polarized} as well as \textit{unpolarized}
particles, in general\@. We have also seen how QED leads directly
to speed dependent entangled states\@. For processes with
initially polarized particles (as well as for spin~$0$ pairs
annihilation into $2\gamma$), a clear violation of Bell's
inequality of LHV theories was obtained for all speeds\@. This
clear violation was also true for several speeds for processes
with initially unpolarized particles, but the tests are more
sensitive on the speed for such processes\@. The main results of
the paper are given in (\ref{Eqn2.20}), (\ref{Eqn2.22}),
(\ref{Eqn2.23}), (\ref{Eqn2.41})--(\ref{Eqn2.43}),
(\ref{Eqn3.10}), (\ref{Eqn3.12})--(\ref{Eqn3.19})\@. We feel that
it is a matter of some urgency that the relevant experiments are
carried out by monitoring speed\@.

\section*{ACKNOWLEDGMENTS}

\noindent\indent The authors would like to acknowledge with thanks
for being granted the ``Royal Golden Jubilee Ph.D. Program'' by
the Thailand Research Fund (Grant No. PHD/0022/2545) for
especially carrying out this project\@. The authors would like
also to thank the referees for for pointing out some pertinent
references for this work and for valuable suggestions\@.


\begin{thebibliography}{99}

\bibitem{B01}
N.~Yongram and E.~B.~Manoukian,
\textit{Int.~J.~Theor.~Phys.}~\textbf{42}, 1755 (2003) or ArXiv: quant-ph/0411072.

\bibitem{B02}
J.~F.~Clauser and M.~A.~Horne, \textit{Phys.~Rev.}~\textbf{D10},
526 (1974).; J.~F.~Clauser and A.~Shimoney,
\textit{Rep.~Prog.~Phys.}~\textbf{41}, 1881 (1978).

\bibitem{B03}
F.~Selleri, in: \textit{Quantum Mechanics Versus Local Realism},
Edited by F.~Selleri (Plenum, New York) 1988.

\bibitem{B04}
A.~Aspect, J.~Dalibard, and G.~Roger,
\textit{Phys.~Rev.~Lett.}~\textbf{49}, 1804 (1982).

\bibitem{B05}
E.~S.~Fry, \textit{Quantum Optics}~\textbf{7} 227 (1995).

\bibitem{B06}
L.~R.~Kaday, J.~D.~Ulman and C.~S.~Wu,
\textit{Nuovo~Cimento}~\textbf{B25}, 633 (1975).


\bibitem{B07}
S.~Osuch, M.~Popskiewicz, Z.~Szeflinski, and Z.~Wilhelmi,
\textit{Acta Phys. Polonica} \textbf{B27}, 567 (1996).

\bibitem{B08}
V.~D.~Irby, \textit{Phys.~Rev.}~\textbf{A67}, 034102 (2003).
ArXiv: quant-ph/0209158.

\bibitem{B09}
A.~Peres and D.~R.~Terno, \textit{Rev.~Mod.~Phys.}~\textbf{76}, 93
(2004). ArXiv: quant-ph/0212023.

\bibitem{B10}
R.~Haag, \textit{Local Quantum Fields, Particles, Algebras}
(Springer, Berlin, 1996).

\bibitem{B11}
H.~Araki, \textit{Mathematical Theory of Quantum Fields} (Oxford,
Oxford, 1999).

\bibitem{B12}
O.~Bratelli and D.~W.~Robinson, \textit{Operator Algebras and
Quantum Statistical Mechanics}, 2nd ed. (Springer, New York,
1987).

\bibitem{B13}
S.~J.~Summers and R.~Werner, \textit{Phys.~Lett.}~\textbf{A110},
257 (1985).; \textit{J.~Math.~Phys.}~\textbf{28}, 2440 (1987).

\bibitem{B14}
L.~J.~Landau, \textit{Phys.~Lett.}~\textbf{A120}, 54 (1987).

\bibitem{B15}
P.~M.~Alsing and G.~J.~Milburn, ArXiv: quant-ph/0203051; R.~M.~Gingrich
and C.~Adami, ArXiv: quant-ph/0205179; S.~D.~Bartlett and D.~R.~Terno,
ArXiv: quant-ph/0403014.

\bibitem{B16}
R.~M.~Gingrich, A.~J.~Bergou and C.~Adami,
\textit{Phys.~Rev.}~\textbf{A68}, 042102 (2003). ArXiv:
quant-ph/0302095.

\bibitem{B17}
D.~R.~Terno, \textit{Phys.~Rev.}~\textbf{A67}, 014102 (2003).
ArXiv: quant-ph/0208074.

\bibitem{B18}
M.~Czachor, \textit{Phys.~Rev.}~\textbf{A55}, 72 (1997). ArXiv:
quant-ph/9609022.

\bibitem{B19}
R.~P.~Feynman, \textit{QED: The Strange Theory of Light and
Matter} (Princeton, New Jersey) 1985.

\bibitem{B20}
J.~Pachos and E.~Solono, ArXiv: quant-ph/0203065.

\bibitem{B21}
C.~Itzykson and J.-B.~Zuber, \textit{Quantum Field Theory}
(McGraw-Hill, New York, 1980).

\end{thebibliography}
\end{document}